# The driving force of labor productivity

Ivan O. Kitov & Oleg I. Kitov

**Abstract**
Labor productivity in developed countries is analyzed and modeled. Modeling is based on our previous finding that the rate of labor force participation is a unique function of GDP per capita. Therefore, labor productivity is fully determined by the rate of economic growth, and thus, is a secondary economic variable.
Initially, we assess a model for the U.S. and then test it using data for Japan, France, the UK, Italy, and Canada. Results obtained for these countries validate those for the U.S. The evolution of labor force productivity is predictable at least at an 11-year horizon.

**Introduction**

Mainstream economics sees labor productivity as the central problem for the understanding of economic evolution. An elevated rate of the growth in labor productivity in the 1990s was considered by Blinder and Yellen (2002) as the driving force of the excellent economic performance. Correspondingly, a slightly lowered growth rate in the 1970s was responsible for "the woeful macroeconomic performance of that decade". Bearing in mind the importance of labor productivity for theoretical and practical purposes, we would like to answer two basic questions:

1. What is the driving force behind the growth of labor productivity?
2. Is it possible to control this force and to achieve stable economic growth?

Quantitative answers to these questions would allow elaboration of a set of reasonable policies in many areas aimed at the acceleration of real economic growth.

From our previous experience with analyzing and modeling real economic growth, among numerous aspects associated with the study of labor productivity, we are specifically interested in its link to the growth in real GDP per capita and to labor force participation rate. For example, Campbell (1994) and Pakko (2002) reported that a decrease in the growth rate of productivity rate results in increasing in employment and output. In several working papers (Kitov, 2006ab; Kitov, Kitov, and Dolinskaya, 2007a) we demonstrated that the evolution of real GDP per capita in the USA is driven by the change in the number of 9-year-olds. In turn, Kitov and Kitov (2008) found that the evolution of labor force participation rate is controlled by real GDP per capita as the only driving force. By definition, labor productivity is a ratio of real GDP and the number of employed persons (or the number of worked hours). Hence, the growth in productivity is also driven by the only macroeconomic variable – real GDP per capita (or the change in the specific age population).

Conventional economics includes extensive literature devoted to the understanding and modeling of the forces (beyond that we investigate here) behind real economic growth. Handbook of Economic Growth (2005) is a valuable source of relevant information and references. Since we are focusing on the aforementioned links in this paper, we explain long- and mid-term trends as well as short-term fluctuations in labor productivity using only real economic growth. Our analysis for the USA is supported and validated by a cross-country comparison.

The remainder of the paper is organized as follows. Section 1 presents some working assumptions and quantitative relationships between labor productivity, labor force participation rate, the growth rate of real GDP per capita, and the number of 9-year-olds. In Section 2, we test these relationships against actual data and present some predictions of the future evolution of productivity in the USA. Section 3 extends our analysis to some other developed countries. Section 4 concludes.

1. **The model**

For the estimation of (average) labor productivity, $P$, one needs to know total output, GDP, and the level of employment, $E$ ($P=GDP/E$), or total number of working hours, $H$ ($P=GDP/H$). First definition includes employment, which is usually determined in the Current Population Surveys conduced at a monthly rate by the U.S. Census Bureau. In the first approximation and for the purposes of our modeling, we neglect the difference between the employment and the level of labor force because the number of unemployed is only a small portion of the labor force. There is no principal difficulty, however, in the subtraction of the unemployment, which is completely defined by the labor force level with possible complication in some countries induced by time lags (Kitov, 2006cd, Kitov, Kitov, and Dolinskaya, 2007b). Hence, a more accurate relationship between productivity and real GDP per capita is potentially available.

The number of working hours is an independent measure of the workforce. Employed people do not have the same amount of working hours. Therefore, the number of working hours may change without any change in the level of employment and vice versa. In this study, the estimates associated with $H$ are used as an independent measure of productivity and for demonstration of the inherent uncertainty in definitions of labor productivity.

Obviously, individual productivity varies in a wide range in developed economies. In order to obtain a hypothetical true value of average labor productivity one needs to sum up individual productivity of each and every employed person with corresponding working time. This definition allows a proper correction when one unit of labor is added or subtracted and distinguishes between two states with the same employment and hours worked but with different productivity. Hence, both standard definitions are slightly biased and represent approximations to the true productivity. Due to the absence of true estimates of labor productivity and related uncertainty in the approximating definitions we do not put severe constraints on the precision in our modeling and seek only for a visual fit between observed and predicted estimates.

Real GDP in the definitions of labor productivity is a measured macroeconomic variable. There is no need to model it in this study and we use the estimates reported by the Bureau of Economic Analysis (BEA). Second term ($E$ or $H$) in the definitions can be and is actually measured. At the same time, it has been definitely driven by one exogenous variable since the mid-1960s. Recently, we developed a model (Kitov and Kitov, 2008) describing the evolution of labor force participation rate, $LFP$, in developed countries as a function of a single defining variable – real GDP per capita, $G$. Natural fluctuations in real economic growth unambiguously lead to relevant changes in labor force participation rate as expressed by the following relationship:

$$\{B_1 dLFP(t)/LFP(t) + C_1\} exp\{\alpha_1[LFP(t) - LFP(t_0)]/LFP(t_0)\} =$$
$$= \int \{dG(t-T))/G(t-T) - A_1/G(t-T)\}dt \qquad (1)$$

where $B_1$ and $C_1$ are empirical (country-specific) calibration constants, $\alpha_1$ is empirical (also country-specific) exponent, $t_0$ is the start year of modeling, $T$ is the time lag, and $dt=t_2-t_1$, $t_1$ and $t_2$ are the start and the end time of the time period for the integration of $g(t) = dG(t-T))/G(t-T) - A_1/G(t-T)$ (one year in our model). Term $A_1/G(t-T)$, where $A_1$ is empirical constant, represents the evolution of potential economic growth (Kitov, 2006b). The exponential term defines the change in the sensitivity to $G$ due to deviation of the $LFP$ from its initial value $LFP(t_0)$. Relationship (1) fully determines the behavior of the $LFP$ when $G$ is an exogenous variable.

It follows from (1) that productivity can be represented as a function of $LFP$ and $G$, $P \sim G \cdot N/N \cdot LFP = G/LFP$, where $N$ is the working age population. Hence, $P$ is a function of $G$ only. Therefore, the growth rate of labor productivity can be represented using several independent variables. Because the change in productivity is synchronized with that in $G$ and labor force participation, first useful form mimics (1):

$$dP(t)/P(t) = \{B_2 dLFP(t)/LFP(t) + C_2\} \cdot exp\{\alpha_1[LFP(t) - LFP(t_0)]/LFP(t_0)\} \qquad (1')$$

where $B_2$ and $C_2$ are empirical calibration constants. Inherently, the participation rate is not the driving force of productivity, but (1′) demonstrates an important feature of the link between $P$ and *LFP* – the same change in the participation rate may result in different changes in the productivity depending on the level of the *LFP*.

In order to obtain a simple functional dependence between *P* and *G* one can use two alternative forms of (1), as proposed by Kitov and Kitov (2008):

$$\{B_3 dLFP(t)/LFP(t) + C_3\} \exp\{\alpha_2[LFP(t) - LFP(t_0)]/LFP(t_0)\} = N_9(t-T)$$
$$dP(t)/P(t) = B_4 N_9(t-T) + C_4 \qquad (2)$$

where $N_9$ is the number of 9-year-olds, $B_3,..., C_4$ are empirical constant different from $B_2$, $C_2$, and $\alpha_2=\alpha_1$. In this representation, we use our finding that the evolution of real GDP per capita is driven by the change rate of the number of 9-year-olds (specific age for U.S. population). Relationship (2) links *dP/P* and $N_9$ directly.

The next relationship defines *dP/P* as a nonlinear function of *G* and serves a workhorse for those countries, which do not provide accurate estimates of specific age population. General nonlinear dependence between *P* and G is as follows:

$$N(t_2) = N(t_1) \cdot \{ 2[dG(t_2-T)/G(t_2-T) - A_2/G(t_2-T)] + 1\} \qquad (3)$$

$$dP(t_2)/P(t_2) = N(t_2-T)/B + C \qquad (4)$$

where *N(t)* is the (formally defined) specific age population, as obtained using $A_2$ instead of $A_1$, *B* and *C* are empirical constants. Relationship (3) defines the evolution of some specific age population, which is different from actual one. This discrete form is useful for calculations.

So, there are three different relationships to test. We use a simplified form of testing procedure - visual fit between measured and predicted rate of productivity growth. The estimates of goodness-of-fit obtained using linear regression analysis are facultative ones.

## 2. Modeling the evolution of productivity in the U.S.

There are several sources of productivity estimates. We use the estimates reported by the Conference Board (2008) and the Bureau of Labor Statistics (2008). Our model predicts the change rate of labor productivity. The upper panel in Figure 1 presents four time series which correspond to two different definitions of productivity. Two curves represent output in U.S. dollars per one hour (ratio of total output and total working hours in the USA). Other two curves represent output per employed person per year. These four curves span the period between 1960 and 2007 and demonstrate similar overall behavior with a deep trough around 1980. Also, notice a decline in productivity since 2003 for all definitions.

Amplitudes of fluctuations clearly differ between the curves. Output per person is characterized by a slightly higher volatility. Due to the observed uncertainty in definitions and measurements one should not expect any model to precisely reproduce these curves. The lower panel depicts the original time series smoothed by a centered 5-year moving average, MA(5). Only output per person estimated by the BLS still has negative values near 1980.

At first, we test our basic hypothesis that the evolution of labor force participation can define that of labor productivity. As discussed above, we replace employment, $E$, in the definition of productivity with *LFP*. Thus, one has to estimate coefficients $B_2$, $C_2$, and $α_1$, which provide the best (visual) fit between the observed and predicted time series. Figure 2 depicts two curves reported by the BLS (both *GDP/E* and *GDP/H*) and those predicted with $B_2$= -5.0, $C_2$=0.040, and $α_1$=5.0; and $B_2$= -3.5, $C_2$=0.042, and $α_1$=3.8, respectively. Due to volatility in the original productivity and labor force (time derivative) series we replace them with their MA(5). A five-year-long time interval provides an increased resolution and allows smoothing measurement noise. As expected, coefficient $B_2$ is negative implying a decline in productivity with increasing labor supply. Both exponents $α_1$ are positive. According to (1′), this fact indicates that the sensitivity of productivity to changes in labor force participation (or to real economic growth) increases with the level of LFP. The goodness-of-fit for both observed time series is about ($R^2$=) 0.6. Moreover, principal features (troughs and peaks) of the observed series are similar in the predicted series, with small time shifts, however. One can approximately divide the whole period into two segments - before and after 1990. In these segments, the predicted curves lag behind and lead the observed ones, respectively.

As in our previous paper (Kitov and Kitov, 2008), the number of 9-year-olds is obtained from the Census Bureau (2008). [Also, we use here the estimates of real GDP per capita (in 1990 U.S. dollars converted at Geary Khamis PPPs) as presented by the Conference Board (2008).] These population estimates allow to model labor productivity as defined by (2). Figure 3 compares the BLS (per hour) labor productivity and that obtained using (2) with the following coefficients: $B_3$=48000000, $C_3$=-0.062, and $T$=2 years. The overall fit is reasonably good with $R^2$=0.46. There is a period of large discrepancy between the observed and predicted curves in the mid-1980s. As with labor force participation, the predicted time series leads the observed one by 2 years.

Figure 4 compares two different predictions of productivity: the one obtained from the *LFP* and that from the $N_9$. The LFP predicted curve is smoothed with MA(5) and that from $N_9$ is obtained with original annual readings. The agreement between these two curves is slightly better than their agreement with the measured productivity curves. This better agreement might be related to the fact that both predictions are associated with population estimates and the productivity is estimated using real GDP reported by the BEA. Revisions to these different time series might be not synchronous and create time shifts in the curves. Problems induced by numerous revisions to population and GDP time series deserve more attention.

Relationship (2) provides a unique opportunity to predict the evolution of productivity at an 11-year-long time horizon. The number of 9-year-olds can be extrapolated 9 years ahead using population estimates in younger cohorts. Additional 2 years are related to the time lag between real economic growth and productivity change. Figure 5 presents population estimates for 9-, 6- and 1-year-olds as published by the U.S. Census Bureau (2008). The curves for younger ages are shifted ahead by 3 and 8 years is order to synchronize them with the current estimates of 9-year-olds. The level of the curves increases with time due to positive overall migration, i.e. the number of 9-year-olds in a given year is the number of 1-year-olds 8 years before plus net migration less total deaths. We are interested in the change rate of $N_9$, however. The lower panel in Figure 5 demonstrates that the estimates of the change rate are very close for all three cohorts, except some short periods, where revisions to these series were different. This closeness implies that one can replace the change rate of $N_9$ 8 years ahead with the current change rate of the number of 1-year-olds. Hence, productivity in the USA will grow at an elevated (relative to potential) rate during the 2010s. This process will be obviously accompanied by an associated decrease in the LFP. If the population estimates are accurate, one can expect sharp changes in real economic growth, labor force participation, and thus, in productivity.

Relationships (3) and (4) define productivity as a function of real economic growth. Figure 6 shows the difference between real GDP per capita and productivity. (Linear regression gives the goodness-of-fit of $R^2=0.61$.) The productivity varies with lower amplitudes because employment is synchronized with the evolution of $G$ and total population does not depend on $G$. Two potential growth rates related to $G$ and $P$ are also shown: $A_1/G$ and $A_2/G$, where $A_1$=$420 and $A_2$=$398. The potential rate of real economic growth is slightly higher than that for productivity. Both constants are determined with high accuracy because even small deviations in the rates results in large deviations in cumulative growth. Therefore, the difference between $A_1$ and $A_2$ is significant despite the curves are so close. The observed productivity curve was below its potential between 1965 and 1982. As compensation, this period is characterized by intensive growth in labor force participation. Since 1982, the productivity fluctuates around its potential.

Results of productivity modeling by (3) and (4) are presented in Figure 7. (Model parameters are given in Figure captions.) Overall, 60% of variability in observed curve is explained by the predicted one – same as explained by *G* itself. Timing of main turns in the curves is excellent. This is an expected effect, however, because productivity is essentially the same class variable as real GDP per capita. An important feature to predict is amplitude, as Figure 6 indicates - productivity is not a scaled version of real GDP per capita. So, the success of our model is related to a good prediction of LFP. Modeling of the evolution of productivity for other developed countries using relevant GDP per capita is necessary to validate our model.

### 3. Modeling labor productivity in developed countries

In this Section, we use only relationship (3) and (4) for the prediction of labor productivity in a number of developed countries. Essentially, these are the countries for which we modeled labor force participation – France, Italy, Canada, Japan, and the UK. These countries are the largest developed economies in the world. The evolution of productivity in Germany is not modeled because of side effects induced by the reunification in the 1990s.

The upper panel in Figure 8 presents observed and predicted productivity growth rate in France. It has been decreasing from 0.05 $y^{-1}$ in the 1960s to 0.02 $y^{-1}$ in the 2000s. This is the result of real economic growth below its (relatively high) potential rate defined by $A_2$=$450, as the lower panel in Figure 8 demonstrates. Both productivity curves are well synchronized but are non-stationary. This might make problematic the results of linear regression analysis with $R^2$=0.91 due to a possibility of spurious regression. However, both variables include real GDP as the main part. Hence, high correlation between them is not a surprise. All in all, the predicted curve is in excellent agreement with the observed one and this observation confirms the results for the USA and supports our model.

Figure 9 depicts observed and predicted productivity for Italy. These curves are similar to those for France and are also in an excellent agreement: the goodness-of-fit is also very high - ($R^2$=) 0.9. The range of productivity change for Italy is even larger: from 0.08 $y^{-1}$ in the 1960s to 0.0 $y^{-1}$ in the 2000s. Hence, real economic growth has been far below its potential rate since 1960s. The current rate of productivity growth is very low and one should not expect any break in the declining trend. The growth rate of labor force has been hovering around zero line since the mid-1970s.

The case of Canada adds some new features to our analysis. Figure 10 displays measured and predicted rate of productivity growth. The curves are very close with $R^2$=0.8, but are characterized by the presence of three peaks – in the early 1960s, between 1983 and 1987, and around 1995. This pattern is quite different from that observed in the USA – the closest neighbor and main trade partner. So, real economic evolution in Canada and U.S. is likely to be independent. For Canada, the range of productivity change is smaller than that in France and Italy: from 0.03 $y^{-1}$ in the 1960s to 0.0 $y^{-1}$ in 1980. The current rate of productivity growth is also close to 0.

United Kingdom and Japan are presented in Figure 11 and 12, respectively. They are similar in sense that accurate prediction from *G* is possible only after 1970. The predicted curves describe amplitudes and timing of major turns in the observed curves. The discrepancy before 1970 is not well explained and might be linked to revisions to employment and real economic growth definitions and measurement errors.

## 4. Conclusion

In Introduction, we formulated two general questions. To answer the first of them, we have modeled productivity growth in the largest (except Germany) developed economies – the USA, Japan, the UK, France, Italy, and Canada. With a varying level of success, the growth rate of productivity is explained by the influence of a single driving force – real GDP per capita. As a dependent variable, labor productivity can be predicted at various time horizons with the uncertainty determined only by the accuracy of population estimates.

The answer to the second question is – "yes". Productivity, as an economic variable, is of a secondary importance. The growth of GDP per capita above or below its potential rate, as defined by the term *$A_2/G$*, is transferred one-to-one in relevant changes in labor force participation and, thus, in employment and productivity. Since real economic growth depends only on the evolution of specific age population, one must control demographic processes in order to control productivity and stable economic growth.

One may also conclude that all attempts to place labor productivity in the center of conventional theories of real economic growth are practically worthless. Productivity is not an independent variable, which can be influenced and controlled by any means except demography.

**Figures**

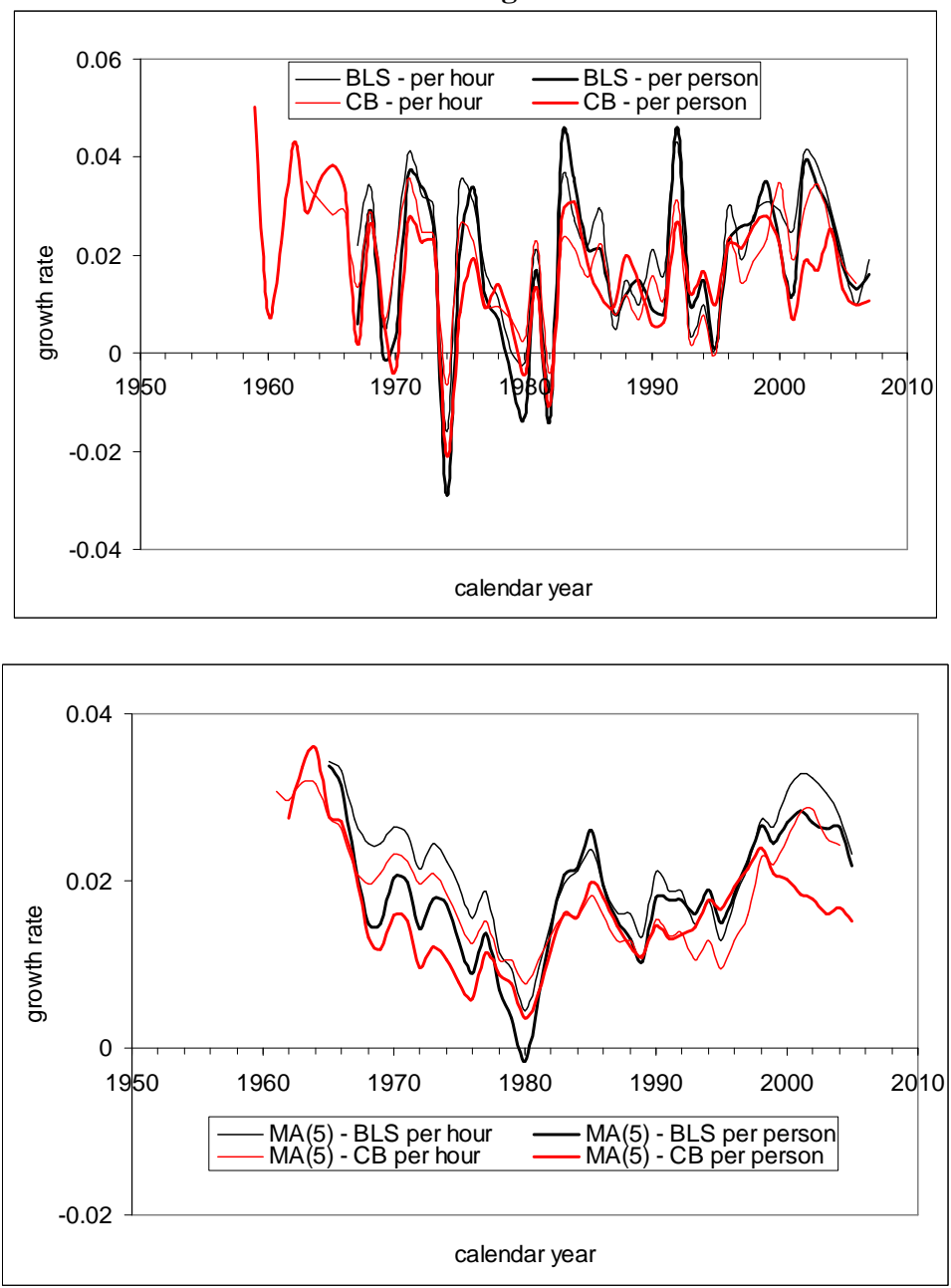

Figure 1. The upper panel displays four time series for two definitions of the growth rate of labor productivity in the USA as expressed in $ per hour and in $ per person per year – for both the Conference Board and the BLS. These series span the period between 1960 and 2007. Output per person is characterized by a slightly higher volatility. Notice the decline in the productivity since 2003 for all three definitions. The lower panel depicts the original time series smoothed by a centered MA(5). Only the BLS measured output per person still has negative values near 1980.

a)

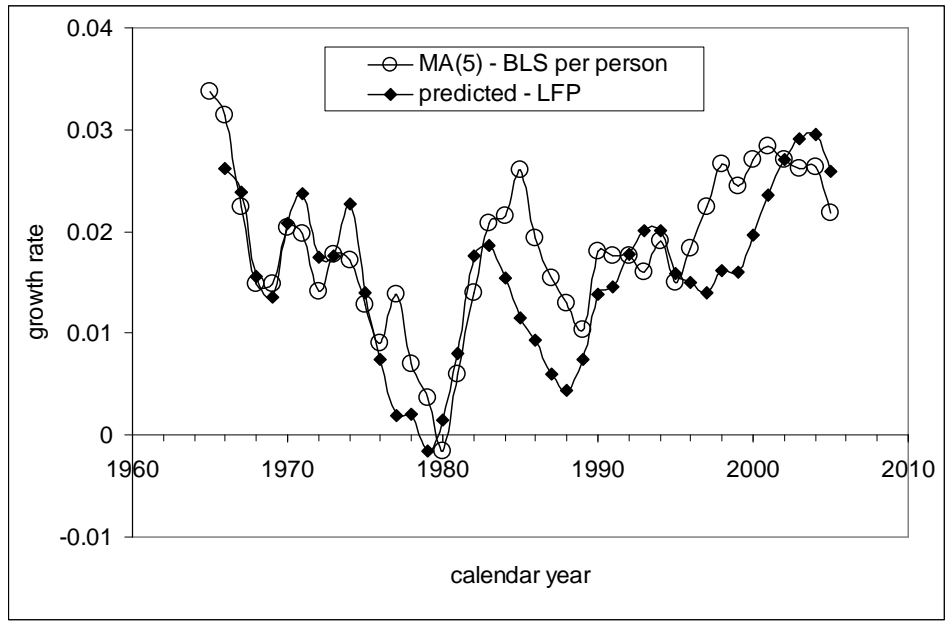

b)

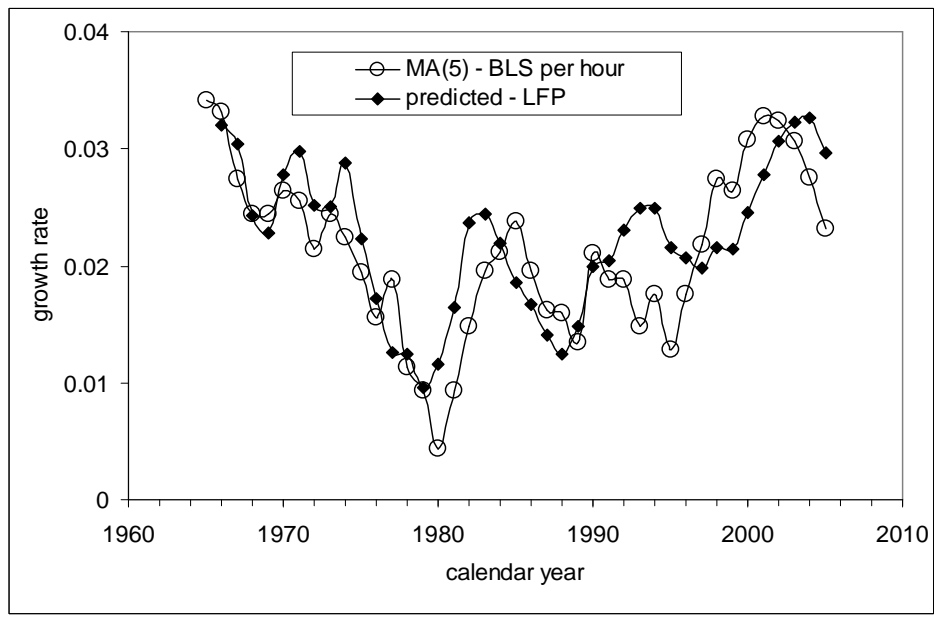

Figure 2. Observed and predicted growth rate of labor productivity. Two BLS measures of productivity are presented: **a)** output ($) per person ; **b)** output ($) per hour.
Linear regression gives close results - $R^2=0.6$ in both cases.

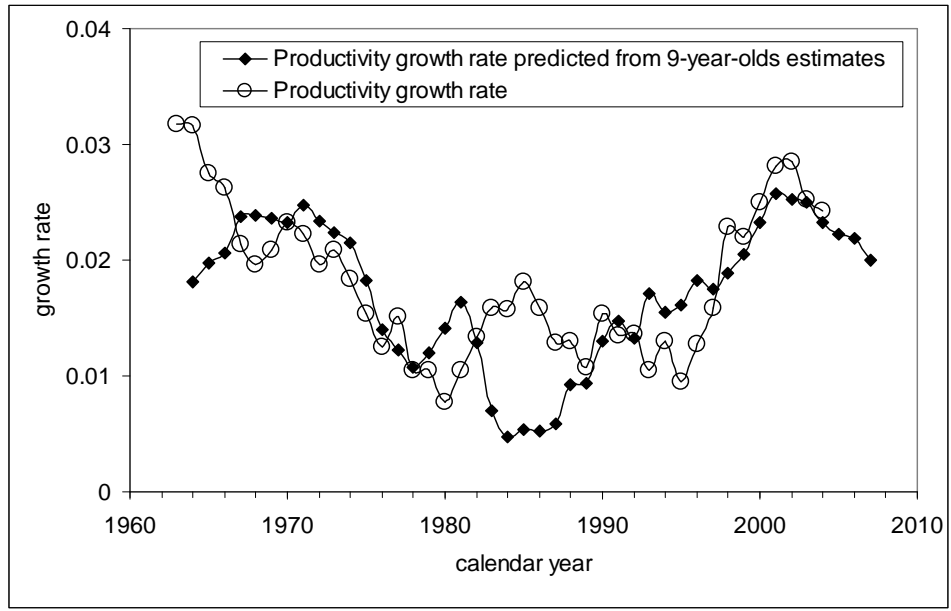

Figure 3. Observed and predicted change rate of productivity (BLS output per hour). The observed curve is represented by MA(5) of the original one. The predicted rate is obtained from the number of 9-year-olds according to relationship (2). Main features of the observed growth are relatively well predicted by the evolution of the number of 9-year-olds. $R^2=0.46$.

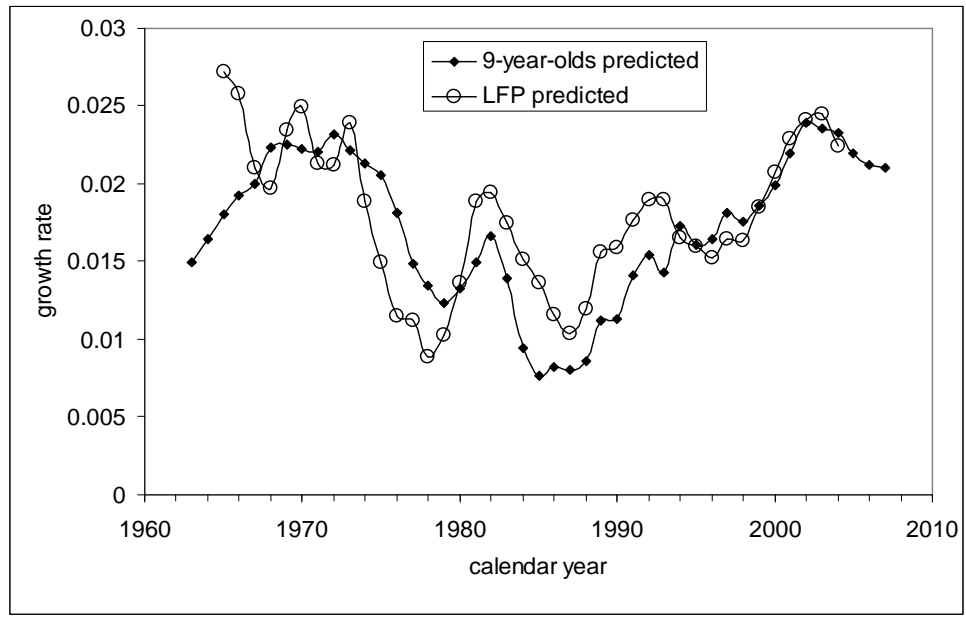

Figure 4. The growth rate of productivity as predicted from the LFP and the number of 9-year-olds. Main fluctuations in the predicted curves are well synchronized.

a)

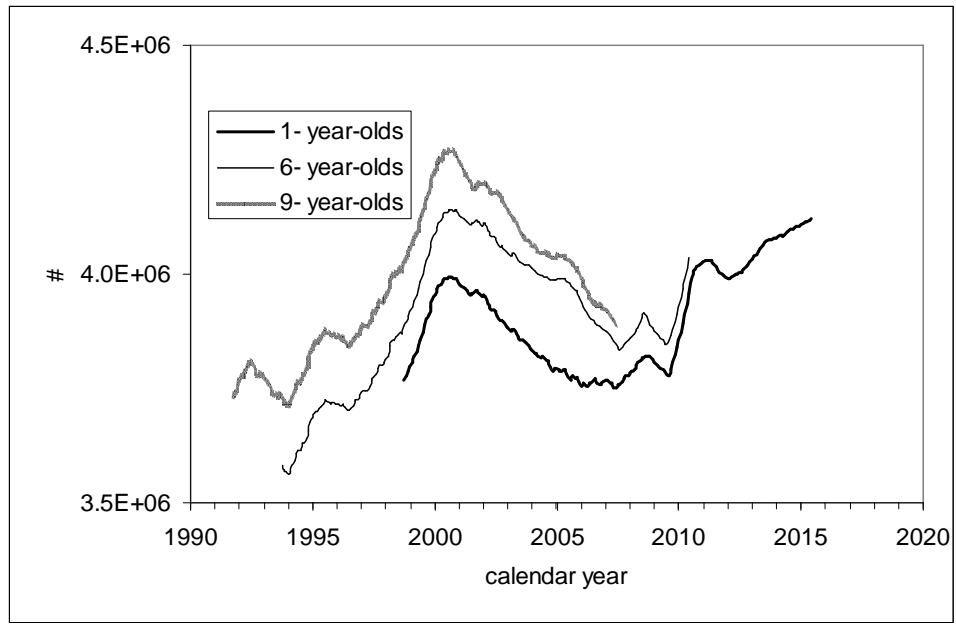

b)

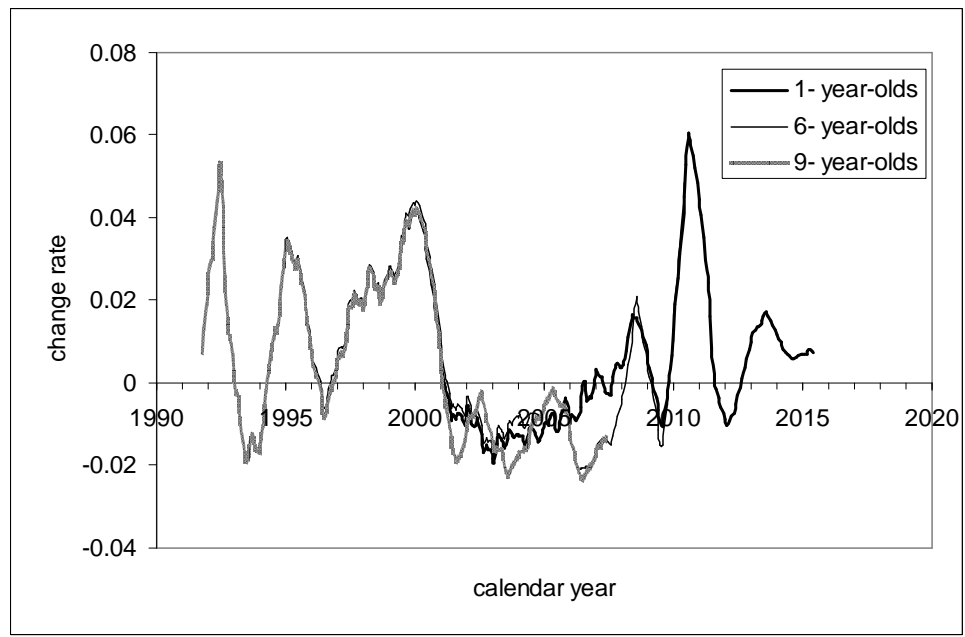

Figure 5. Prediction of the number of 9-year-olds by extrapolation of population estimates for younger ages (1- and 6-year-olds).
**a)** Total population estimates. The time series for younger ages are shifted ahead by 8 and 3 years, respectively.
**b)** Change rate of the population estimates, which is proportional to the growth rate of real GDP per capita. Notice the difference in the change rate provided by 1-year-olds and 6-year-olds for the period between 2003 and 2010. This discrepancy is related to the age-dependent difference in population revisions.

     A downward trend in productivity, as has been observed since 2003, will turn to an upward one in the 2010s. This also means an elevated growth rate of real GDP per capita during the period between 2010 and 2017.

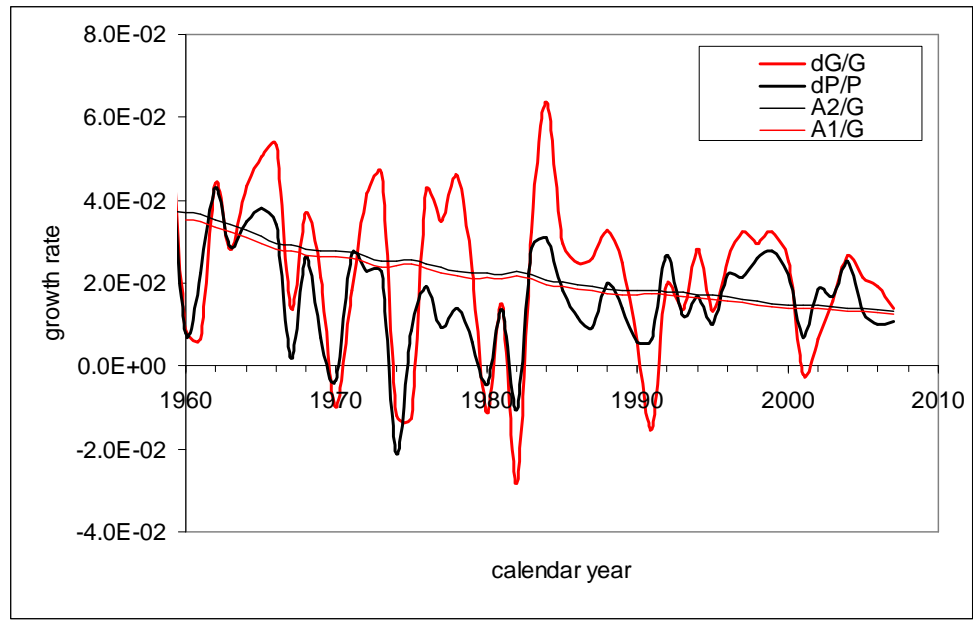

Figure 6. The growth rate of productivity, *dP/P*, and real GDP per capita, *dG/G*. Corresponding potential growth rates $A_1/G$ and $A_2/G$ are also shown. The curve *dP/P* is below the potential between 1965 and 1983.

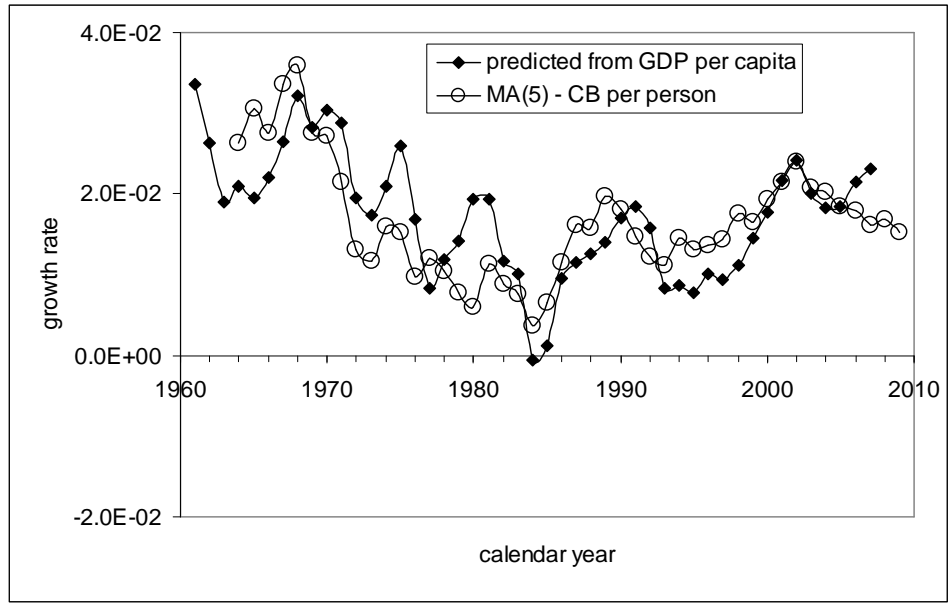

Figure 7. Observed and predicted change rate of productivity (Conference Board GDP per person employed). The observed curve is represented by MA(5) of the original one. Linear regression gives $R^2$=0.6.
Model parameters are as follows: $N(1959)$ = 4500000, $A_2$=\$420, $B$=3500000, $C$=-0.095.

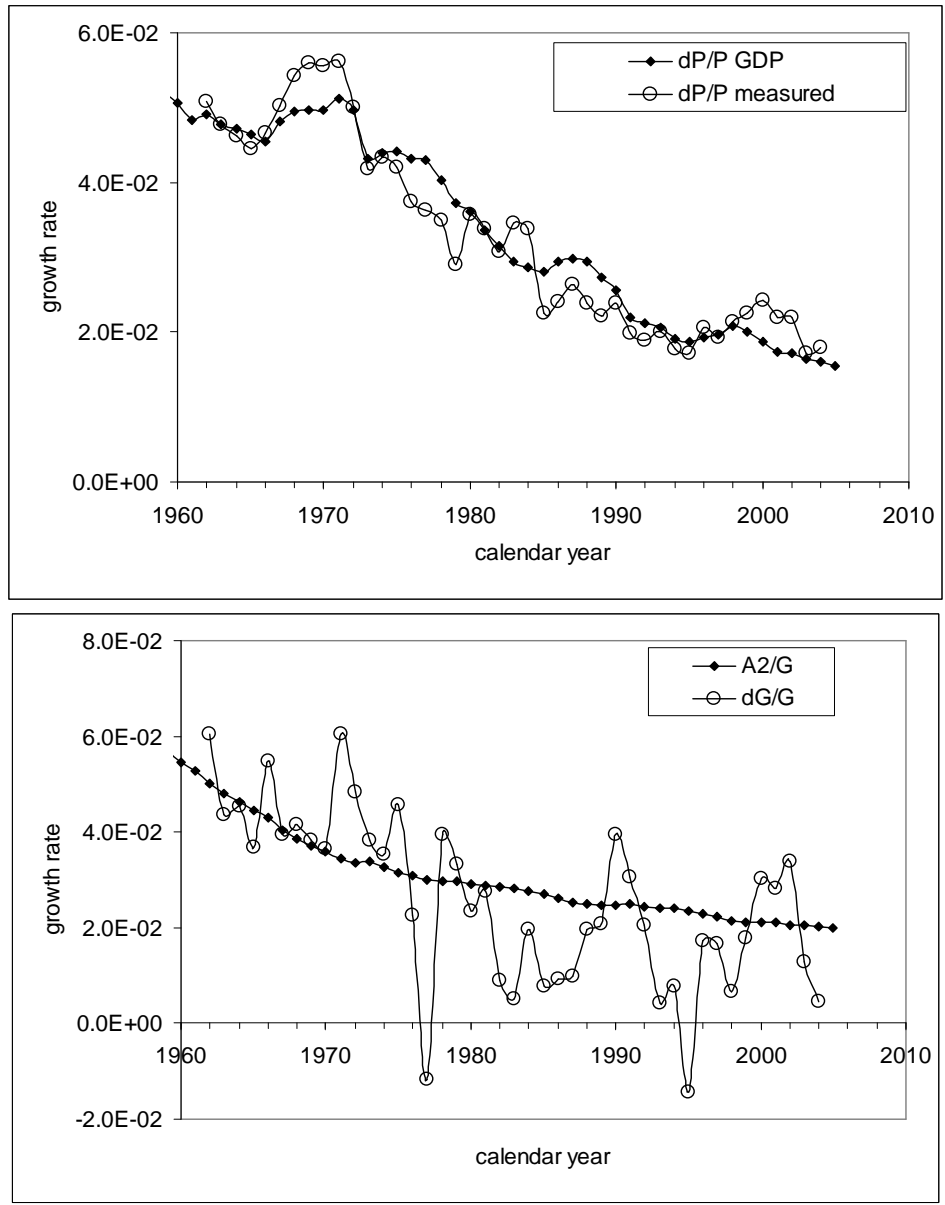

Figure 8. Upper panel: observed and predicted productivity in France. Model parameters: $N(1959)$=570000, $A_2$=\$450, $B$=7500000, $C$=-0.022. $R^2$=0.91. Lower panel: $dG/G$ and $A_2/G$.

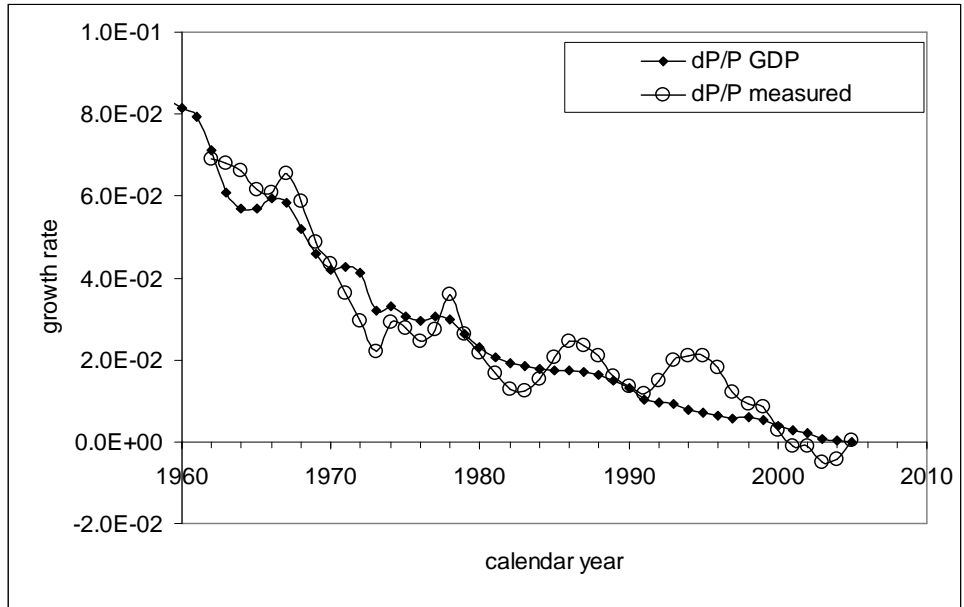

Figure 9. Observed and predicted productivity in Italy: *N(1959)*=570000, $A_2$=\$550, *B*=5000000, *C*=-0.018. $R^2$=0.9.

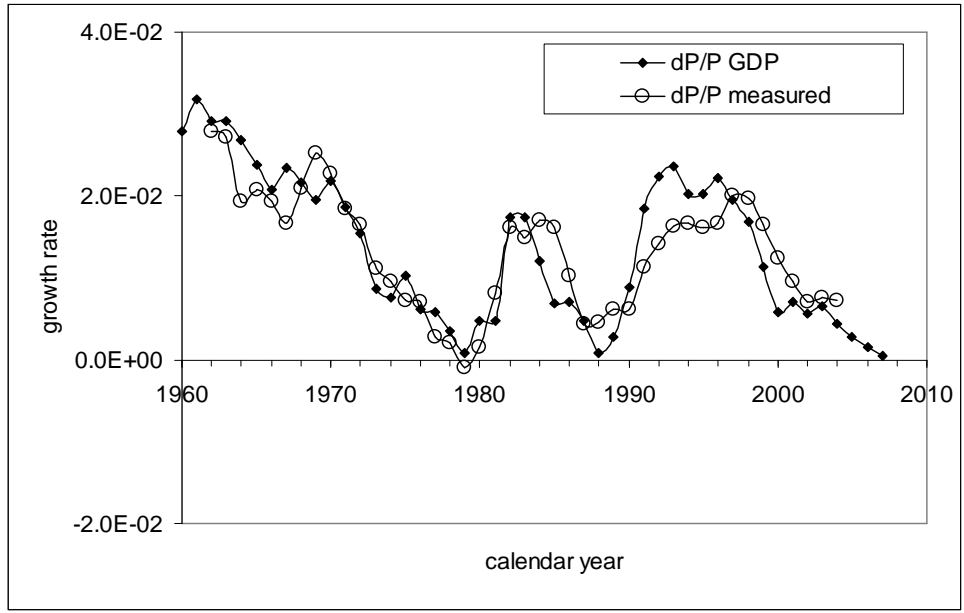

Figure 10. Observed and predicted productivity in Canada: $N(1959)$=270000, $A_2$=\$300, $B$=-3200000, $C$=0.108. $R^2$=0.8.

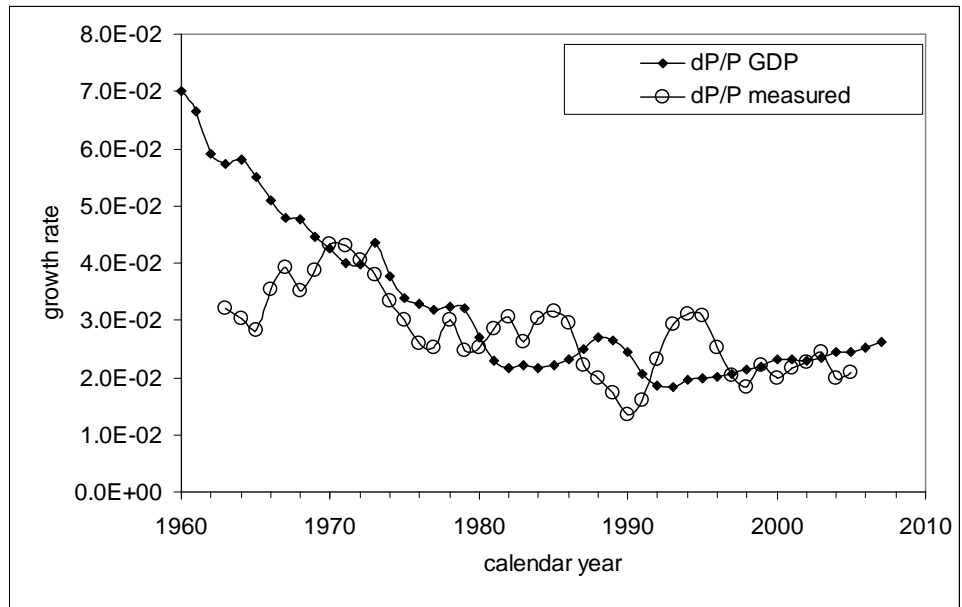

Figure 11. Observed and predicted productivity in the UK: *N(1959)* =670000, $A_2$=$390, *B*=7500000, *C*=-0.02.

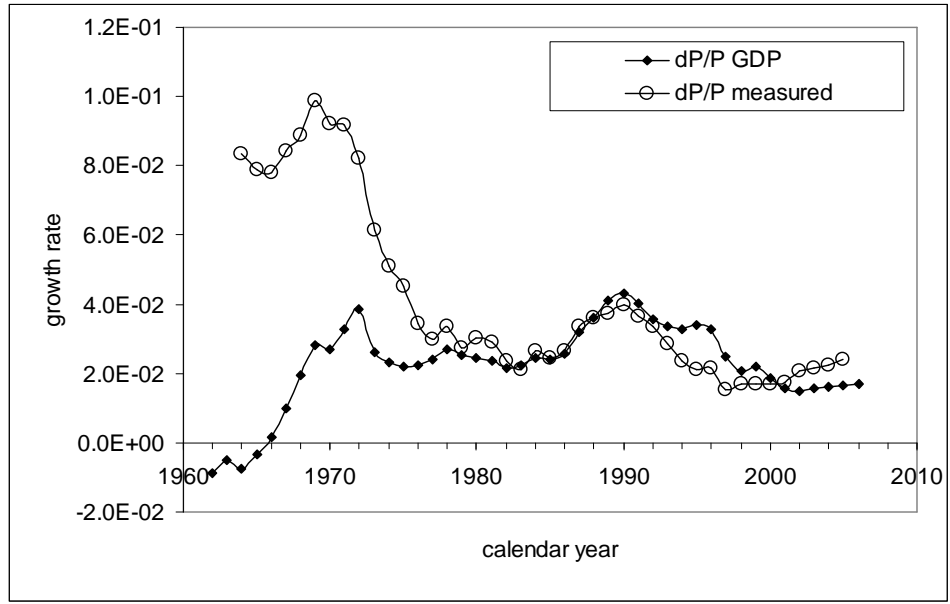

Figure 12. Observed and predicted productivity in Japan: *N(1959)*=2000000, $A_2$=$400, *B*=4000000, *C*=-0.018.